\documentclass{article}

\usepackage{arxiv}

\usepackage[utf8]{inputenc} 
\usepackage[T1]{fontenc}    
\usepackage{url}            
\usepackage{booktabs}       
\usepackage{amsfonts}       
\usepackage{nicefrac}       
\usepackage{microtype}      
\usepackage{lipsum}		
\usepackage{doi}

\usepackage{color,array,amsthm}
\usepackage{graphicx}
\usepackage{amsmath,amsthm}
\usepackage{lineno,hyperref}
\usepackage{longtable,tabularx}
\usepackage[version=4]{mhchem}
\usepackage{siunitx}
\usepackage{float}
\usepackage{dsfont}

\newcommand{\angvel}{\boldsymbol\omega}
\newcommand{\linvel}{V}
\newcommand{\genvel}{\boldsymbol{v}}

\newtheorem{rem}{Remark}

\title{Sequential Data-Assisted Control in Flight}

\author{ \href{https://orcid.org/0000-0003-3344-5068}{\includegraphics[scale=0.06]{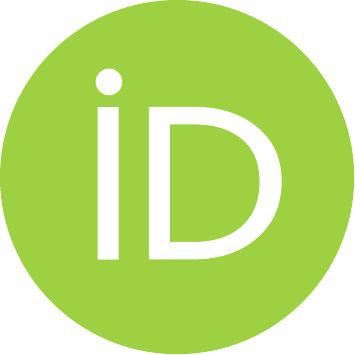}\hspace{1mm}Mostafa Eslami} \\
	Department of Aerospace Engineering\\
	Sharif University of Technology\\
	Azadi Avenue, Tehran \\
	\texttt{eslami.mostafa@ae.sharif.edu} \\
	\And
	\href{https://orcid.org/0000-0000-0000-0000}{\includegraphics[scale=0.06]{orcid.pdf}\hspace{1mm}
 Afshin Banazadeh} \\
	Department of Aerospace Engineering\\
	Sharif University of Technology\\
	Azadi Avenue, Tehran \\
	\texttt{banazadeh@sharif.edu} \\
}

\renewcommand{\shorttitle}{\textit{arXiv} Template}

\hypersetup{
pdftitle={Sequential Data-Assisted Control in Flight},
pdfsubject={Systems and Control},
pdfauthor={Mostafa Eslami, Afshin Banazadeh},
pdfkeywords={Sequential Data-Assisted Control, GTM},
}

\begin{document}
\maketitle

\begin{abstract}
Flight dynamics involve uncertainties in parameters, aerodynamic derivatives, and engine thrust. These uncertainties can be categorized into three types: known-predictable, known-unpredictable, and unknown. While advanced control systems typically rely on high-fidelity dynamical models in dealing with known-predictable uncertainties, simplified approaches are used for the second and third categories to manage the complexities involved in synthesis and implementation. 
In this paper, the focus is on accurately modeling the internal dynamics, which primarily deal with parametric uncertainties. Real-time data is employed to identify uncertainties in the remaining external dynamics, including both known-unpredictable and unknown aspects. To address these uncertainties and maintain optimal performance, stability, and robustness, the authors propose a framework known as Sequential Data-Assisted Control (SDAC). This framework involves using a model-based nonlinear controller for the internal dynamics to provide the desired momentum to a data-based controller responsible for the external dynamics.
By the momentum through the internal dynamics and leveraging the Koopman operator, the linear evolution of momentum is derived. This information is then utilized by the data-based controller to assign appropriate control inputs. The proposed approach establishes a novel foundation for a comprehensive analysis of maneuverability, stabilizability, and controllability.
To evaluate the performance of SDAC, closed-loop simulations are conducted using the NASA Generic Transport Model (GTM). The data-based controller employs Linear Quadratic Regulator (LQR), while the model-based controller uses robust sliding mode control. A comparison is made with a pure robust nonlinear controller, demonstrating significant performance improvements, particularly in cases involving known-unpredictable uncertainties.
\end{abstract}


\section{Introduction}

In the previous work, we proposed a framework for Data-Assisted Control (DAC) for aerospace vehicles, which involves using data to support a model-based controller in extending the performance of the system over a damage event \cite{eslami2023data}. The framework required real-time decisions to override the control law with information obtained from the data, while the model-based controller does not perform perfectly. The NASA Generic Transport Model (GTM) was used as the platform for the study  \cite{jordan2004development,bacon2007general,hueschen2011development}, and simulations have shown that purely model-based robust control leads to degradation of the closed-loop performance in case of damage, suggesting the need for data assistance. To implement the DAC framework, the paper proposed the use of the Dual Unscented Kalman Filter (DUKF) to estimate the fixed dynamic parameters and the Koopman estimator to estimate the evolution of the generalized force moments \cite{koopman1931hamiltonian}, called pseudo-observations. The closed-loop system is shown to be stable under certain attainable conditions in the transition phase between the data and the model. 

However, the paper also notes that the model-based paradigm works optimally for a certain region with known uncertain boundaries and has proven results in abundant practices. The paper suggests that with the assistance of data, control systems can extend their capabilities beyond the limits of the model-based paradigm, but data cannot completely replace model-based methods and theories. A comprehensive review of data-based algorithms in control comparative to model-based control and adaptive laws was presented \cite{eslami2023data}.


As the reader can find, the DAC framework relied on prior knowledge of uncertain terms participating in external dynamics. The Sequential DAC (SDAC) framework follows a similar approach by decoupling exact and uncertain parts and addressing control objectives differently for uncertain parts. In SDAC, momentum dynamics evolve linearly in real-time, with the Koopman operator processing momentum observations to create the linear model. The robust nonlinear controller exploits exact dynamics to generate reference momentum, while the LQR controller employs the linear model to produce deviations in control effectors to attain the desired momentum. This sequential or cascade approach enables concurrent analysis of maneuverability and controllability and eliminates the need for prior knowledge of uncertain terms, although this information may be used in a generalized observable space (lifted-space) to increase the performance. Same as previous work, the GTM is the platform for this study. In the case of parametric uncertainty, the same as DAC, DUKF can be used for the generation of pseudo-observation. For simplicity, it is assumed there are no parametric uncertainties in internal dynamics. 

In recent years, the Koopman operator and linearization successfully have been applied in many aerial and space applications \cite{hofmann2022advances,servadio2022dynamics, arnas2021approximate}. In this paper, it is proved the momentum linearly evolves in time, and the Koopman operator is a promising approach to identifying the linear model of external dynamics through time by data. Since the linear model ensembles momentum as states and flight control effectors as a control input, the maneuverability and controllability of the linear system with state-input constraints would be of great importance.

The SDAC incorporates several novel features, including the removal of the pseudo-inverse of the control input matrix for computing control inputs (control allocation) \cite{durham2017aircraft}, a controllability guarantee for the momentum dynamics assuming that the reference momentum is within the attainable momentum subset of the linear model, the presence of independent control parameters in the robust nonlinear controller of the internal dynamics to ensure attainable momentum, and the absence of a requirement for prior knowledge of uncertain terms. Additionally, there is no switching between two controllers, and after a brief period, the SDAC will use the full capacity of the estimated dynamics, negating the need for a decision-making algorithm.  Furthermore, SDAC enables the potential use of alternative control systems synthesis and analysis methods and algorithms in nature. Finally, compared to DAC, the SDAC has fewer parameters to tune and provides more physical intuition since the dynamics decoupling is over internal and external forces to ensure steady momentum at equilibrium. 

The structure of the paper is outlined as follows: Section \ref{sec:seq_DAC} provides a detailed explanation of the SDAC framework, introducing the fundamental components of the framework. In Section \ref{sec:dynamics_decouple}, the decoupling of internal and external dynamics is discussed, along with an explanation of the model derived from the GTM and external dynamics. Section \ref{sec:control_system} presents the design of two separate controllers for the internal and external dynamics. The analysis of controllability, stabilizability, robustness, and maneuverability of the SDAC is carried out in Section \ref{sec:analysis}. Finally, a series of simulation results is presented in Section \ref{sec:simulations}.

\section*{Nomenclature}
{\renewcommand\arraystretch{1.0}
\begin{table}[H]
\begin{tabular}{|p{0.1\linewidth} p{0.9\linewidth}|}
\hline\hline
$\angvel$  & [rad/sec] angular velocity vector with $p$-roll rate, $q$-pitch rate and $r$-yaw rate arguments in body coordinate \\\cline{2-2}
$\rho$ &  [ft]  center of mass displacement \\\cline{2-2}
$\linvel$ & [ft/sec] linear velocity vector with $u$, $v$ and $w$ arguments in body coordinate\\\cline{2-2}
$\genvel$ & generalized body coordinate velocity vector - $[\linvel^T,\angvel^T]^T$\\\cline{2-2}
$m$ & [lbs] aircraft mass \\\cline{2-2}
$I_M$ & [slug.ft$^2$] aircraft moment of inertia matrix with $I_{ab}$ entries for axis $a$ and $b$\\\cline{2-2}
$I$ & identity matrix with proper dimension\\\cline{2-2}
$g$   & [ft/sec$^2$] gravity constant \\\cline{2-2}
$\eta_1$ & position vector with $X$, $Y$ and $Z$ arguments in earth coordinate\\\cline{2-2}
$\eta_2$ & orientation vector (Euler angles) with $\Phi$, $\Theta$ and $\Psi$ arguments in earth coordinate \\\cline{2-2}
$\eta$ & generalized earth-fixed coordinate position and orientation - $[\eta_1^T,\eta_2^T]^T$ \\\cline{2-2}
$\mathrm{p}$ & parameters set\\\cline{2-2}
$W$ & [lbf] gravitational force \\\cline{2-2}
$F$ & [lbf] external force applied to aircraft \\\cline{2-2}
$T$, $T_R$, $T_L$ & [lbf] total thrusts generated by engines, right engine thrust, left engine thrust \\\cline{2-2}
$M$ & [lbf.ft] external moment applied to aircraft \\\cline{2-2}
$C_x$ & aerodynamic dimensionless coefficients corresponds to force/moment $x$\\\cline{2-2}
$\tau$ & generalized force and moment vector\\\cline{2-2}
$L$ & generalized momentum vector consisting of linear and angular momentums\\\cline{2-2}
$\mathcal{M}$ & mass-inertia matrix\\\cline{2-2}
$\mathcal{C}$ & Coriolis and centrifugal matrix\\\cline{2-2}
$\mathcal{G}$ & gravitational force and moments vector\\\cline{2-2}
$B$ & control derivatives in force and moments\\\cline{2-2}
$D$ & damping derivatives in force and moments\\\cline{2-2}
$C$ & dimensionless aerodynamics coefficient\\\cline{2-2}
$E$ & extra order dynamics\\\cline{2-2}
$c_g$ & center of gravity\\\cline{2-2}
$c_p$ & center of pressure\\\cline{2-2}
$e_x$ & indication of error associated with the variable $x$\\\cline{2-2}
$\omega_x$ & noise vector associated with variable $x$ \\\cline{2-2}
$\sim\mathcal{N}(0,A)$ & zero-mean Gaussian distribution with covariance matrix $A$ \\\cline{2-2}
$\Sigma \cdot$   & summation operator - means total forces or moments \\\cline{2-2}
$\mathcal{S}(\cdot)$ & skew-symmetric operator acts on the vector $(\cdot)$ and generates associated skew-symmetric matrix\\\cline{2-2}
diag$(\cdot)$ & a function generating diagonal matrix of vector $(\cdot)$\\\cline{2-2}
$\|\cdot\|_2$ or $\|\cdot\|$ & norm-2 of vector $(\cdot)$ or Frobenius norm of matrix $(\cdot)$ \\\cline{2-2}
$\underline{\lambda}(\cdot)$, $\overline{\lambda}(\cdot)$ 	& minimum and maximum eigenvalue of matrix $(\cdot)$, respectively\\\cline{2-2}
$\sigma(A)$, $\sigma_i(A)$	& all singular values and $i$-th singular value of matrix $A$\\\cline{2-2}
$(\cdot)^T$	& transpose operator\\\cline{2-2}
$\hat{(\cdot)}$ & estimated parameter, vector or matrix\\\cline{2-2}
$\tilde{(\cdot)}$ & difference between the actual and the desired variable\\\cline{2-2}
$\Breve{(.)}$ & observation or direct measured value\\\cline{2-2}
$\Breve{\mathcal{Y}}$ & observables matrix\\\cline{2-2}
$\lambda$, $\lambda^\ast$, $\lambda_{sel}$ & decision factor, optimum decision factor, and selected decision factor by pilot\\\cline{2-2}
Subscripts & \\
$T$ & Refer to the thrusters\\\cline{2-2}
$A$ & Refer to the aerodynamics\\\cline{2-2}
$F$ & Refer to the forces\\\cline{2-2}
$M$ & Refer to the moments\\\cline{2-2}
$\delta$ & Refer to the control surfaces/actuators\\\cline{2-2}
$er$, $el$, $e$ & right elevator, left elevator, and coordinated left-right elevator together\\\cline{2-2}
$ru$ & rudder\\\cline{2-2}
$ar$, $al$, $a$ & right aileron, left aileron, and coordinated left-right aileron together\\\cline{2-2}
$r$ & reference/residual value\\\cline{2-2}
$a$ & under state variable means augmentation\\\cline{2-2}
$d_i$ & indication of GTM damage case $i$\\\cline{2-2}
$d$ & indication of desired value\\\hline\hline
\end{tabular}
\end{table}}

\section{Sequential DAC}\label{sec:seq_DAC}
\begin{figure}[H]
  \centering
  \includegraphics[width=0.8\columnwidth]{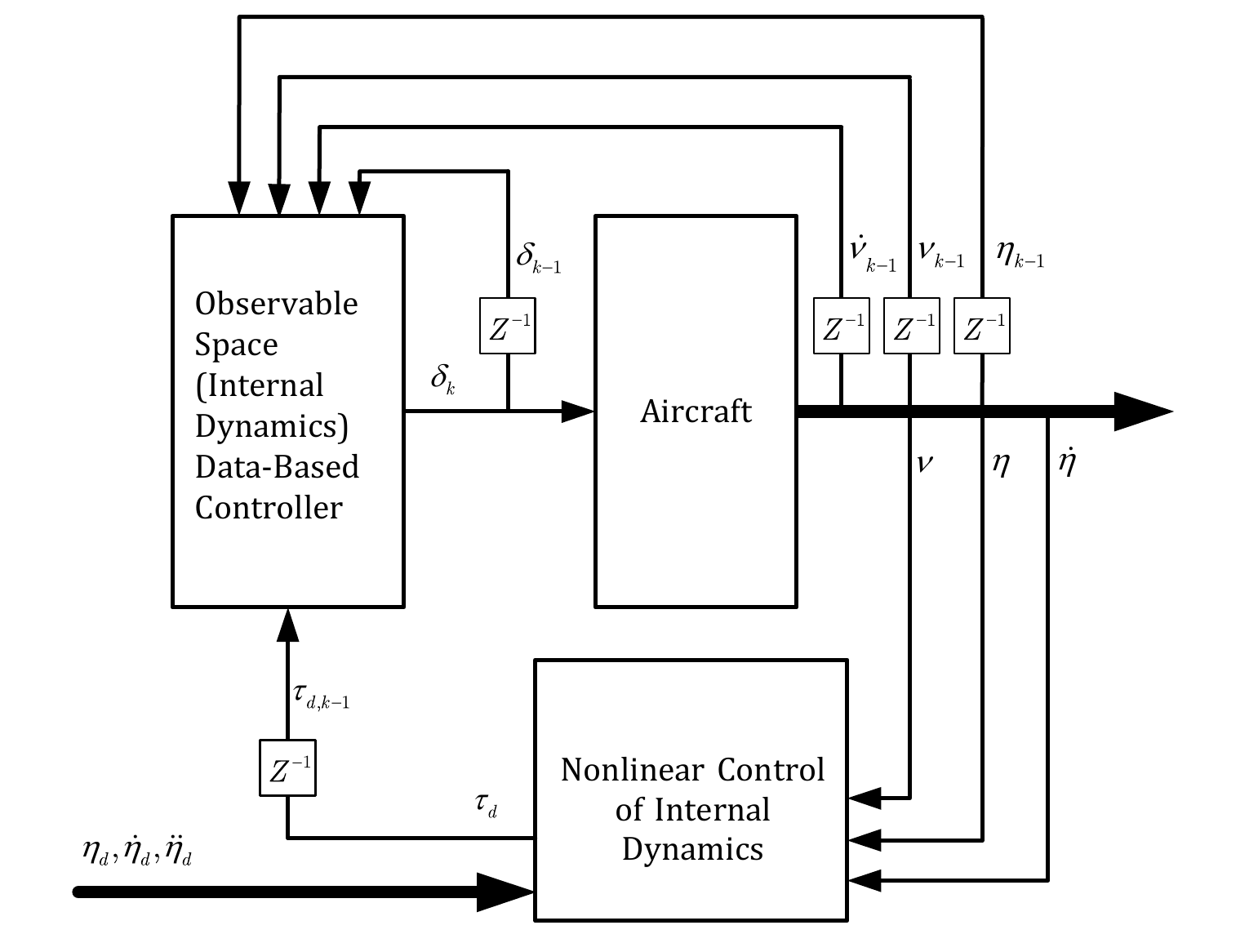}
  \caption{Sequential Data Assisted Control (SDAC) framework for Aircraft}\label{fig:SDAC_nonlinear_GTM}
\end{figure}

The block diagram of the SDAC framework is illustrated in Figure \ref{fig:SDAC_nonlinear_GTM}. It showcases two cascade controllers:
\begin{itemize}
\item A nonlinear controller for the fixed (internal) dynamics, responsible for generating reference momentum to achieve the desired trajectory and velocities.
\item A data-based controller for the uncertain (external) dynamics, estimated in real-time, which achieves the reference momentum by commanding the control inputs of the aircraft.
\end{itemize}

The term "framework" implies that SDAC can be expanded using various methods in nonlinear control of internal dynamics, data-based control of external dynamics, and identification of observable space evolution over time. In this paper, the following approaches are employed:

\begin{itemize}
    \item For the internal dynamics, a robust sliding mode controller is developed.
    \item The observable dynamics are identified using the Koopman operator.
    \item A Linear Quadratic Regulator (LQR) is devised to control the derived linear dynamics obtained from the Koopman operator.
\end{itemize}

\section{Dynamics Decoupling in SDAC}\label{sec:dynamics_decouple}
This section begins by introducing the GTM model, followed by the decoupling of the internal and external dynamics, which is suitable for control system design.

\subsection{GTM Nonlinear Model}\label{sec:GTM}
The set of flight dynamics equations for GTM considering the change in the body's center of mass is described in matrix form as follows \cite{bacon2007general,guo2011multivariable,eslami2023data}. 


\begin{align}\label{equ:eom_closed}
  \mathcal{M}\dot{\genvel}+\mathcal{C}(\genvel)\genvel = \tau_{TA\delta}-\mathcal{G}(\eta) = \tau
\end{align}

where $\tau_{TA\delta}$ is the generalized force and moment generated by the aircraft engine and the aerodynamics of the aircraft and control surfaces.   $\mathcal{M}$ is the mass-inertia matrix, $\mathcal{C}$ is Coriolis and centrifugal matrix, and $\mathcal{G}$ is gravitational force and moment acting on the flight as, 
\begin{align}
  \mathcal{M} = \left[\begin{array}{cc}
               mI & -m\mathcal{S}(\rho)\\
               m\mathcal{S}(\rho) & I_M\\
             \end{array}
           \right], \quad \mathcal{C}(\genvel)=\left[\begin{array}{cc}
               m\mathcal{S}(\angvel) & -m\mathcal{S}(\mathcal{S}(\angvel)\rho)\\
               -m\mathcal{S}(\mathcal{S}(\angvel)\rho) & -\mathcal{S}(I_M\angvel)+m\mathcal{S}(\mathcal{S}(\linvel)\rho)\\
             \end{array}
           \right]
\end{align}

and
\begin{align}
\mathcal{G}(\eta) = \left[\begin{array}{c}
               -W\\
               -\mathcal{S}(\rho)W\\
             \end{array}
           \right], \quad W = \left[
             \begin{array}{c}
               -mg\sin \Theta \\
               mg\cos \Theta \sin \Phi \\
               mg \cos \Theta \cos \Phi \\
             \end{array}
           \right].
\end{align}

\begin{rem}\label{rem:mdot_2c}
  Assuming small change rate in mass and moment of inertia, $\mathcal{M}$ and $\mathcal{C}$ are rearranged such that $\dot{\mathcal{M}}-2\mathcal{C}(\genvel)$ is skew-symmetric. 
\end{rem}

The generalized force and moment correspond to the body's coordinate linear and angular momentum as follows,

\begin{align}
    \tau = \dot{L}+\mathcal{S}^\ast(\angvel)L, \quad \mathcal{S}^\ast(\angvel) = \left[\begin{array}{cc}
        \mathcal{S}(\angvel) & \emptyset  \\
         \emptyset & \mathcal{S}(\angvel)
    \end{array}\right].
\end{align}

Considering the assumptions in the derived model \cite{eslami2023data}, the final form of the nonlinear model will be summarized as,
\begin{align}\label{equ:eom_closed3}
  \mathcal{M}(\mathrm{p})\dot{\genvel}+\mathcal{C}(\mathrm{p},\genvel)\genvel = \tau = \tau_0 -  \mathcal{G}(\mathrm{p},\eta)+ D\genvel  +B\delta + \sum_{i} f_i(\mathrm{p},\genvel,\delta)   .
\end{align}
Where $\sum_{i} f_i(\mathrm{p},\genvel,\delta)$  represents additive uncertainties. In this equation, the control inputs are as follows,
\begin{align}
  \delta = \left[\begin{array}{c}
               \delta_T\\
               \delta_{ru}\\
               \delta_{a}\\
               \delta_{e}\\
             \end{array}
           \right]
\end{align}

This yields coupled generalized velocity and momentum as follows,
\begin{align}
    &\dot{\genvel}=\mathcal{M}^{-1}(\mathrm{p})(D\genvel+\sum_{i} f_i(\mathrm{p},\genvel,\delta)-\mathcal{C}(\mathrm{p},\genvel)\genvel) + \mathcal{M}^{-1}(\mathrm{p})(\tau_0-\mathcal{G}(\mathrm{p},\eta))   +\mathcal{M}^{-1}(\mathrm{p})(B\delta ) \nonumber\\
    &\dot{L} = D\genvel+\sum_{i} f_i(\mathrm{p},\genvel,\delta)-\mathcal{S}^\ast(\angvel)L + \tau_0+B\delta
\end{align}
The linearized model around the equilibrium can be written as,
\begin{align}
\left[\begin{array}{c}
     \dot{\Delta\genvel}  \\
     \dot{\Delta L} 
\end{array}\right] = \left[\begin{array}{cc}
     A_{\genvel_0} & \emptyset \\
     A_{L\genvel_0} & A_{L_0} 
\end{array}\right]\left[\begin{array}{c}
     \Delta\genvel  \\
     \Delta L 
\end{array}\right] + \left[\begin{array}{c}
     B_{\genvel_0}  \\
     B_{L_0} 
\end{array}\right]\Delta\delta
\end{align}
Where $\Delta\genvel=\genvel-\genvel_0$, $\Delta L=L-L_0$ and $\Delta\delta=\delta-\delta_0$, and the jacobians are calculated as follows,

\begin{align}
    A_{\genvel_0} =\mathcal{M}^{-1}(\mathrm{p})\left(D-\dfrac{\partial(\mathcal{C}(\mathrm{p},\genvel)\genvel-\sum_{i} f_i(\mathrm{p},\genvel,\delta))}{\partial \genvel}\right|_{\genvel=\genvel_0},\quad B_{\genvel_0} = \mathcal{M}^{-1}(\mathrm{p})\left(B+\dfrac{\partial}{\partial\delta}\sum_{i} f_i(\mathrm{p},\genvel,\delta)\right|_{\delta=\delta_0}
\end{align}
\begin{align}
    A_{L\genvel_0} =\left( D-\dfrac{\partial(\mathcal{S}^\ast(\angvel)L-\sum_{i} f_i(\mathrm{p},\genvel,\delta)) }{\partial \genvel}\right|_{\genvel=\genvel_0},\quad  A_{L_0} = -\mathcal{S}^\ast(\angvel_0),\quad B_{L_0} = \left. B+\dfrac{\partial}{\partial\delta}\sum_{i} f_i(\mathrm{p},\genvel,\delta)\right|_{\delta=\delta_0}
\end{align}

These equations give trivial result: $\dot{\Delta L} = \mathcal{M}(\mathrm{p})\dot{\Delta\genvel}$, because at equilibrium (trim condition) we have $\angvel_0=0$, $\rho=0$, and $L_0 = \mathcal{M}(\mathrm{p})\genvel_0$., then 
\begin{align}
    \left(\dfrac{\partial\mathcal{S}^\ast(\angvel)L }{\partial \genvel}\right|_{\genvel=\genvel_0} = \left(\dfrac{\partial\mathcal{C}(\genvel)\genvel}{\partial \genvel}\right|_{\genvel=\genvel_0}\quad \&\quad -\mathcal{S}^\ast(\angvel_0) =\emptyset
\end{align} 

using Equation \eqref{equ:momentum} we have Equation-of-Momentum,
\begin{align}\label{equ:momentum}
     \dot{\Delta L} = (A_{L\genvel_0}\mathcal{M}^{-1}(\mathrm{p})+A_{L_0})\Delta L + B_{L_0}\Delta \delta
\end{align}
\begin{rem}
    Assuming a piece-wise change in the parameters, the linear model is still valid around the initial value of $\genvel_{0_i}$ in the $i$-th piece. Because the parameter set will be fixed in the window time of each piece and we can write    $\left.\dot{\Delta L}\right|_{\genvel_{0_i}} = \left.\mathcal{M}(\mathrm{p})\dot{\Delta\genvel}\right|_{\genvel_{0_i}}$.
\end{rem}

\subsection{Evolution of Momentum Using Koopman Operator}
The left-hand side of equation \ref{equ:eom_closed3}, fixed-dynamics, collects the set of dynamics related to the directly measured variables, i.e. $\dot{\genvel}$ and $\genvel$. However, the right-hand side includes the forces and moments with a range of unknown to known dynamics that evolve in time. Observation of the momentum is directly calculated via available measurements as follows,

\begin{align}
    \Breve{L}(t) = \int_{-\infty}^{t}\mathcal{M}(\mathrm{p}_0)\Breve{\dot{\genvel}}+\mathcal{C}(\mathrm{p}_0,\Breve{\genvel})\Breve{\genvel}-\mathcal{S}^\ast(\Breve{\angvel})\Breve{L}(\tau)\mathrm{d}\tau
\end{align}

these equations imply correct observation of momentum only realizes when parameters are known or its estimation is in hand. So-called pseudo-observation of $L$ in DAC,
\begin{align}
    \Breve{L}(\hat{\mathrm{p}},t) = \int_{-\infty}^{t}\mathcal{M}(\hat{\mathrm{p}})\Breve{\dot{\genvel}}+\mathcal{C}(\hat{\mathrm{p}},\Breve{\genvel})\Breve{\genvel}-\mathcal{S}^\ast(\Breve{\angvel})\Breve{L}(\hat{\mathrm{p}},\tau)\mathrm{d}\tau
\end{align}


 According to Koopman approach, the evolution of the observables, $g(x)$, can be described with a linear system,
\begin{align}\label{equ:koopman_linear}
    x_{k+1} = Ax_k+Bu_k
\end{align}
and, exploiting the Dynamic Mode Decomposition with control approach, called DMDc \cite{proctor2016dynamic}, the couple $(A,B)$ are identified. The approach is as follows:
\begin{align}
    X=\left[\begin{array}{c|c|c|c}
         x_1 &  x_2 & \ldots & x_m\\
    \end{array}\right], \quad X'=\left[\begin{array}{c|c|c|c}
         x_2 &  x_3 & \ldots & x_{m+1}\\
    \end{array}\right]\quad \text{and}\quad U = \left[\begin{array}{c|c|c|c}
         u_1 &  u_2 & \ldots & u_m\\
    \end{array}\right].
\end{align}
Therefore the linear system may be written as,
\begin{align}
    X'=AX+BU = \left[\begin{array}{cc}
        A & B \\
    \end{array}\right]\left[\begin{array}{c}
        X  \\
        U\\
    \end{array}\right] =G\Omega
\end{align}
The $G$ matrix is obtained via least-squares regression:
\begin{align}
    G\approx X'\Omega^{\dagger}
\end{align}
The matrix $\Omega$ is a high-dimensional data matrix. In order to avoid computational errors, the DMDc approach suggests a similar approach to DMD using SVD of $\Omega$. This approach yields reduced space of eigenvectors associated with nonzero eigenvalues of $\Omega$.

Since the state equation reveals one the trivial basis of state evolution is the states themselves, and since $x$ cannot participate as observables in vector field $g(x)$ \cite{brunton2016koopman}, the $g(x)=x=\Delta L$ is an obvious choice. It is worth noting the matrix A in \eqref{equ:koopman_linear} closely resembles the Koopman operator \cite{rowley2009spectral}. 

\section{Control System}\label{sec:control_system}
In this section, the design of the data-based controller for the momentum dynamics using LQR and the model-based robust sliding mode controller for the internal dynamics is presented.

\subsection{LQR for External Dynamics}
The identified data-driven linear dynamics in the previous section are engaged in linear optimal control problem to yield $\tau\rightarrow\tau_d$, i.e.
\begin{align}\label{equ:momentumtozero}
    \Delta L \rightarrow 0, \quad L\rightarrow L_d\quad\text{and}\quad\dot{L}\rightarrow\dot{L}_d\;.
\end{align}
The $L_d$ and $\dot{L}_d$ are defined so that $\genvel\rightarrow\genvel_d$ with governing of fixed-dynamics with robust nonlinear control.

The LQR controller control law for observable space can be written as \cite{kirk2004optimal},
\begin{align}
  u^\ast = -R^{-1}B^TKx
\end{align}
where the cost function is defined as,
\begin{align}
  J_c = \dfrac{1}{2}\tilde{x}^T(t_f)H\tilde{x}(t_f)+\dfrac{1}{2}\int_{t_0}^{t_f} \left(\tilde{x}^TQ\tilde{x} + u^TRu\right)\mathrm{d}t
\end{align}
and the state feedback gain, $K$, can be calculated by solving the Ricatti equation,
\begin{align}
  \dot{K} = -KA-A^TK-Q+KBR^{-1}B^TK
\end{align}

The gain $K$ obtained from this Riccati equation ensures that the Lyapunov function $V=0.5\tilde{x}^TP_1\tilde{x}$, where $P_1$ is positive definite, exhibits a negative derivative over time equal to $\dot{V}=-0.5\tilde{x}^TQ\tilde{x}$, indicating asymptotic stability.



\subsection{Robust Sliding-Mode Controller for Fixed Dynamics}\label{sec:control_law}
Equation \eqref{equ:eom_closed3} ensembles nonlinear dynamics in body coordinates. The objective of the control system design is tracking in earth and body coordinates.  Accordingly, the dynamics together with Kinematic constraints form the earth-fixed dynamics as follows. In the next, it will be seen that the reformulation guarantees condition $\genvel\rightarrow\genvel_d$ and $\eta\rightarrow\eta_d$.
\begin{align}
     &\mathcal{M}\dot{\genvel}+\mathcal{C}(\genvel)\genvel = \tau\nonumber\\
    &\dot\eta=J(\eta_2){\genvel}
\end{align}
 $J(\eta_2)$ is the Euler Transformation (ET) matrix which brings the inertia frame into alignment with the body-fixed frame and $J^{-1}(\eta_2)$  is the Inverse Euler Transformation (IET) matrix which brings the body-fixed frame into alignment with the inertial frame,
\begin{align}
J(\eta_2)=\begin{bmatrix}
            J_1(\eta_2) & 0 \\
            0 & J_2(\eta_2)\\
          \end{bmatrix}\quad\&\quad J^{-1}(\eta_2)=\begin{bmatrix}
            J_1^{-1}(\eta_2) & 0 \\
            0 & J_2^{-1}(\eta_2)\\
          \end{bmatrix}
\end{align}
and,
\begin{align*}
&J_1(\eta_2)=\begin{bmatrix}
              c(\Psi)c(\Theta) & -s(\Psi)c(\Phi)+c(\Psi)s(\Theta)s(\Phi) & s(\Psi)s(\Phi)+c(\Psi)c(\Phi)s(\Theta) \\
              s(\Psi)c(\Theta) & c(\Psi)c(\Phi)+s(\Phi)s(\Theta)s(\Psi) & -c(\Psi)s(\Phi)+s(\Theta)s(\Psi)c(\Phi) \\
              -s(\Theta) & c(\Theta)s(\Phi) & c(\Theta)c(\Phi) \\
            \end{bmatrix}\\&J_1^{-1}(\eta_2)=J_1^T(\eta_2)\\
&J_2(\eta_2)=\begin{bmatrix}
              1 & s(\Phi)t(\Theta) & c(\Phi)t(\Theta) \\
              0 & c(\Phi) & -s(\Phi) \\
              0 & \dfrac{s(\Phi)}{c(\Theta)} & \dfrac{c(\Phi)}{c(\Theta)} \\
            \end{bmatrix}\;\&\;J_2^{-1}(\eta_2)=\begin{bmatrix}
              1 & 0 & -s(\Theta) \\
              0 & c(\Phi) & c(\Theta)s(\Phi) \\
              0 & -s(\Phi) & c(\Theta)c(\Phi) \\
            \end{bmatrix}\\
\end{align*}
where $s(.) = \sin(.)$, $c(.) = \cos(.)$ and $t(.) = \tan(.)$. 
The dynamics in the earth frame coordinates can be written as follows:
\begin{align}\label{equ:rov_earthfixed}
    \mathcal{M}_\eta(\eta)\ddot{\eta}+ \mathcal{C}_\eta(\eta,\genvel)\dot{\eta}=\tau_\eta
\end{align}
where,
\begin{align*}\label{equ:dynaimc_trans}
   & \mathcal{M}_\eta(\eta)=J^{-T}(\eta)\mathcal{M}J^{-1}(\eta)\\
   & \mathcal{C}_\eta(\eta,\genvel)=J^{-T}(\eta)\left(\mathcal{C}(\genvel)-\mathcal{M}J^{-1}(\eta)\dot{J}(\eta)\right)J^{-1}(\eta)\\
   & \tau_\eta=J^{-T}(\eta)\tau.
\end{align*}
translating the dynamic model in earth frame preserves the skew-symmetric property of  $\dot{\mathcal{M}}_\eta(\eta)-2\mathcal{C}_\eta(\eta,\genvel)$, since $\mathrm{d}({J}^{-1}(\eta))/\mathrm{d}t=J^{-1}(\eta)\dot{J}(\eta)J^{-1}(\eta)$, and,
\begin{align}
  \dot{\mathcal{M}}_\eta(\eta) = -2\dot{J}^{-T}(\eta)\mathcal{M}J^{-1}(\eta)\dot{J}(\eta)J^{-1}(\eta)\Rightarrow 
   \dot{\mathcal{M}}_\eta(\eta)- 2\mathcal{C}_\eta(\eta,\genvel)= J^{-T}(\eta)\left(\mathcal{M}-2\mathcal{C}(\genvel)\right)J^{-1}(\eta)
\end{align}

Assuming the model parameters are known (i.e. ($\mathcal{M}$ and $\mathcal{C}(\genvel)$), and the values of $\eta$ , $\dot{\eta}$ are available, the control law becomes \cite{gao.xue.2006}\cite{garcia.salgado.torres.2009}\cite{uzmay.burkan.sarikaya.2004},
\begin{align}\label{equ:control_law}
    {\tau}=J^T(\eta)(\mathcal{M}_{\eta}(\eta)\dot{{\pmb{\vartheta}}}+\mathcal{C}_{\eta}(\eta,{\genvel}){\pmb{\vartheta}} -\Gamma \pmb{s})
\end{align}
where,
\begin{align}\label{equ:slotine.li.1991_exchange}
{\pmb{\vartheta}}=\dot{\eta}_d-\Lambda\tilde{\eta}\;\text{ \& }\;\pmb{s}=\dot{\eta}-{\pmb{\vartheta}}
\end{align}
and $\tilde{\eta}=\eta-\eta_d$ is the tracking error, and $\Lambda$ and $\Gamma$ are diagonal positive definite matrices (gain matrices) defined as,
\begin{align}
\Lambda=\text{diag}([\lambda_1 \;\lambda_2 \;... \;\lambda_6])\;\text{and}\;\Gamma=\text{diag}([\gamma_1 \;\gamma_2 \;... \;\gamma_6])
\end{align}
Therefore, the closed-loop becomes,
\begin{align}
&\mathcal{M}_{\eta}(\eta)(\ddot{\eta}-\dot{{\pmb{\vartheta}}})+\mathcal{C}_{\eta}(\eta,{\genvel})(\dot{\eta}-{\pmb{\vartheta}})+\Gamma \pmb{s} =0\; \Rightarrow\;\\
&\mathcal{M}_{\eta}(\eta)\dot{\pmb{s}}+\mathcal{C}_{\eta}(\eta,{\genvel})\pmb{s}+\Gamma \pmb{s}=0\nonumber
\end{align}
which describes a nonlinear ordinary differential equation with $\pmb{s}$ as the variable. To determine closed-loop stability $V=0.5 \pmb{s}^T \mathcal{M}_{\eta}(\eta) \pmb{s}$ is chosen as the candidate Lyapunov function. Since $\dot{\mathcal{M}}_{\eta}(\eta)-2\mathcal{C}_{\eta}(\eta,{\genvel})$ is skew symmetric matrix, this leads to,
\begin{align}\label{equ:stability}
    \dot{V}=-\pmb{s}^T \Gamma \pmb{s} \leq 0 \Rightarrow \pmb{s}\rightarrow 0 \Rightarrow \tilde{\eta}\rightarrow 0 \Rightarrow \eta\rightarrow \eta_d \text{ and } \dot{\tilde{\eta}}\rightarrow 0 \Rightarrow \dot{\eta}\rightarrow\dot{\eta}_d
\end{align}
Therfore,
\begin{align}\label{equ:norm_ineq}
        \underline{\lambda}(\Gamma)\|\pmb{s}\|^2 \leq \pmb{s}^T \Gamma \pmb{s} \leq \overline{\lambda}(\Gamma)\|\pmb{s}\|^2 \Rightarrow \dot{V}\leq -\underline{\lambda}(\Gamma)\|\pmb{s}\|^2
\end{align}
and since  $V=0.5 \pmb{s}^T \mathcal{M}_{\eta}(\eta) \pmb{s}$, then,
\begin{align}\label{equ:norm_equ_v}
        \frac{1}{2}\underline{\lambda}(\mathcal{M}_{\eta}(\eta))\|\pmb{s}\|^2 \leq \frac{1}{2}\pmb{s}^T \mathcal{M}_{\eta}(\eta) \pmb{s} \leq \frac{1}{2}\overline{\lambda}(\mathcal{M}_{\eta}(\eta))\|\pmb{s}\|^2
\end{align}
From \eqref{equ:norm_ineq} and \eqref{equ:norm_equ_v} one may conclude that,
\begin{align}\label{equ:covergence_v}
    \dot{V}\leq -2\left(\frac{\underline{\lambda}(\Gamma)}{\overline{\lambda}(\mathcal{M})}\right)V \Rightarrow V(t)\leq V(0)e^{-2\left(\dfrac{\underline{\lambda}(\Gamma)}{\overline{\lambda}(\mathcal{M})}\right)t}
\end{align}
Where $V(0)=0.5 \pmb{s}(0)^T \mathcal{M}_{\eta}(\eta(0)) \pmb{s}(0)\geq0$. The role of $\underline{\lambda}(\Gamma)$ now becomes evident. By increasing its value, $V(t)$ can be made to converge to zero exponentially faster. Moreover, the eigenvalues $\lambda_i$ control the speed at which the error is eliminated $\tilde{\eta}\rightarrow 0$ \cite{eslami2019robust}.

The derived model for the GTM is expected to have some discrepancy with the actual dynamics in data-driven dynamics. Moreover, the GTM is subject to considerable environmental disturbances and external forces, parametric changes, and measurement errors. To study the effects of these disturbances, let ${\pmb{d}(t)}$ be an extra force-moment vector in \ref{equ:rov_earthfixed} which denotes the sum of all possible external disturbances. As with the study of uncertain linear systems, the upper bound of ${\pmb{d}(t)}$, i.e. $\|{\pmb{d}(t)}\| \leq \chi$ is considered known. 
By adding an extra term $\pmb{u}_0\in \mathds{R}^6$, the stability of the closed-loop system will be guaranteed in the presence of these uncertainties\cite{sato.tsurta.mukai.2007}\cite{liu.goldenberg.1994}\cite{effatnejad.namvar.2009}. Consider the equations of motion, subject to the uncertainty described above,
\begin{align}
\mathcal{M}_\eta(\eta)\ddot{\eta}+\mathcal{C}_{\eta}(\eta,\genvel)\dot{\eta}=\tau_{\eta}+{\pmb{d}(t)}
\end{align}
Using \eqref{equ:slotine.li.1991_exchange} the closed-loop equation become,
\begin{align*}
\mathcal{M}_\eta(\eta)\dot{\pmb{s}}+\mathcal{C}_{\eta}(\eta,\genvel)\pmb{s}+\Gamma \pmb{s}={\pmb{d}(t)}
\end{align*}
addition of $\pmb{u}_0$ to the control law leads to,
\begin{align*}
\mathcal{M}_\eta(\eta)\dot{\pmb{s}}+\mathcal{C}_{\eta}(\eta,\genvel)\pmb{s}+\Gamma \pmb{s}={\pmb{d}(t)}+\pmb{u}_0
\end{align*}
To determine closed-loop stability, take $V=0.5\pmb{s}^T\mathcal{M}\pmb{s}$ to be the candidate Lyapunov function,
\begin{align*}
\dot{V} = -\pmb{s}^T\Gamma \pmb{s} +\pmb{s}^T {\pmb{d}(t)}+\pmb{s}^T \pmb{u}_0
\end{align*}
Now, by letting $\pmb{u}_0 = -\chi \dfrac{\pmb{s}}{\|\pmb{s}\|}$,
\begin{align}
\dot{V} \leq -\underline{\lambda}(\Gamma)\|\pmb{s}\|^2 +\chi\|\pmb{s}\|+\pmb{s}^T \pmb{u}_0 =  -\underline{\lambda}(\Gamma)\|\pmb{s}\|^2 \leq 0 \Rightarrow \pmb{s}\rightarrow 0 \nonumber
\end{align}
Clearly $\|\pmb{u}_0\|$ is bounded in $[-\chi,\;+\chi]$, however as $\pmb{s}$ goes through zero, chattering will ensue. To eliminate chattering the additional term is modified as follows \cite{khalil.2002},
\begin{align}\label{equ:robust}
\pmb{u}_0=-\chi\tanh{\left(\dfrac{\pmb{s}}{\varepsilon}\right)}
\end{align}

\section{Analysis}\label{sec:analysis}
In this section, we will analyze the controllability, stabilizability, robustness, and maneuverability of SDAC in a systematic manner.

\subsection{Controlability}
The internal dynamics under robust nonlinear control law \eqref{equ:control_law} is always controllable iff $\dot{\mathcal{M}}-2\mathcal{C}(\genvel)$ is skew-symmetric. Because the closed-loop system poles are selected independently for each dimension and the convergence rate reaching the local space can be selected independently for each dimension. Therefore, in the final time from any initial state, the closed-loop system can reach to the final state.

Controllability of the external dynamics under LQR controller can be assessed directly with couple $(A,B)$. In which, $(A,B)$ are controllable iff the controllability matrix $\mathbb{C}(A,B)$ is full rank,
\begin{align}
\mathbb{C}(A,B) = \left[B\;AB\;\ldots\;A^5B \right]
\end{align}

\subsection{Stabilizability}
The stability of the robust nonlinear controller closed-loop system is guaranteed iff $\dot{\mathcal{M}}-2\mathcal{C}(\genvel)$ is skew-symmetric. The stability of the LQR closed-loop system is guaranteed iff uncontrollable modes of the couple $(A,B)$ are stable. If couple $(A,B)$ are controllable, there is state feedback gain $K$ that stabilizes the momentum dynamics. 

Let the Lyapunov function be defined as follows,
\begin{align}
    V = \dfrac{1}{2}\pmb{s}^T\mathcal{M}\pmb{s}+\dfrac{1}{2}\tilde{\tau}^T P_2\tilde{\tau}
\end{align}
Negative definiteness of $\dot{V}$ ensures the asymptotic stability  of SDAC. Let assume $P_2=(A-BK)^{-T}P_1(A-BK)^{-1}$, then,
\begin{align}
    \dot{V} =-\pmb{s}^T\Gamma\pmb{s}+\dfrac{1}{2}\dot{\tilde{\tau}}^T (A-BK)^{-T}P_1(A-BK)^{-1}\tilde{\tau}+\dfrac{1}{2}\tilde{\tau}^T (A-BK)^{-T}P_1(A-BK)^{-1}\dot{\tilde{\tau}}
\end{align}

From momentum equation we have $\tilde{\tau} = \dot{\tilde{L}}$ and $\dot{\tilde{\tau}} = \ddot{\tilde{L}}$. Also from LQR design we have $\dot{\tilde{L}}=(A-BK)\tilde{L}$ and $\ddot{\tilde{L}}=(A-BK)\dot{\tilde{L}}$. Therefore,
\begin{align}
    \dot{V} =-\pmb{s}^T\Gamma\pmb{s}+\dfrac{1}{2}\dot{\tilde{L}}^TP_1\tilde{L}+\dfrac{1}{2}\tilde{L}^TP_1\dot{\tilde{L}}
\end{align}
If we select $P_1$ the same positive definite matrix in LQR Lyapunov function, $\dot{\tilde{L}}^TP_1\tilde{L}+\tilde{L}^TP_1\dot{\tilde{L}}$ is always negative definite. As a direct conclusion, SDAC is stabilizable if the couple $(A,B)$ is controllable.

\begin{rem}
    $P_2=(A-BK)^{-T}P_1(A-BK)^{-1}$ is  positive definite. Because $P_1$ is positive definite, thus there is a matrix $A$ in which $P_1=A^TA$. Therefore $P_2$ can be decomposed as $P_2=X^TX$ where $X=P_1(A-BK)^{-1}$. Since $(A-BK)$ is non-singular, matrix $X$ always exists and accordingly, $P_2$ is positive definite.
\end{rem}

\subsection{Robustness}
Proof of SDAC robustness with the same approach of using $\pmb{u}_0$ in \eqref{equ:robust} is straightforward.

\subsection{Manoeuvrability}
We can denote the maximum attainable momentum subset as $\mathbb{M}$, the admissible control inputs as $\mathbb{U}$, and the attainable momentum subset for admissible control input as $\mathbb{M}^+$. It is always true that $\mathbb{M}^+\subseteq \mathbb{M}$. For a flight to be considered maneuverable, the generated time history of momentums $L(t)$ for any $u\in\mathbb{U}$ must fall into $\mathbb{M}^+$, denoted as $L(t)\in\mathbb{M}^+$ for $t\geq 0$ \cite{durham2017aircraft}.

The SDAC framework allows for maneuverability assessment without the need for a control allocation algorithm or dynamic inversions. In SDAC, for a reachable final momentum in momentum dynamics, the reference momentum generated by the robust-nonlinear controller must be a member of $\mathbb{M}^+$. Therefore, if $L_d(t)\in\mathbb{M}^+$ for $t\geq 0$, the flight is considered maneuverable. Corrective action, such as mitigation on maneuver demands, can then be carried out to transfer reachable reference momentum to the LQR controller.

Assessing maneuverability becomes challenging in cases of uncertainty. It is impossible to estimate $\mathbb{M}^+$ for all cases of uncertainties. However, SDAC provides a rigorous approach to solve this problem using the concept of reachability in linear systems analysis and synthesis. From this perspective, a flight is considered maneuverable only if $L_d\in\mathbb{M}$, only if the couple of $(A,B)$ are stabilizable, and if the momentum dynamics are reachable for constrained inputs. Although reachability with input and state constraints is an ongoing research topic in control, there are elaborated methods for practical cases \cite{d1992reachability,alanwar2022robust,bak2021reachability}. From a synthesis point of view, the LQR design with constraints in the inputs and states can overcome the complexity of control allocation in cases of uncertainty \cite{lee2022gpu,mare2007solution,johansen2000explicit,scokaert1998constrained}.



\section{Simulations}\label{sec:simulations}
\begin{figure}[H]
  \centering
  \includegraphics[width=0.9\columnwidth]{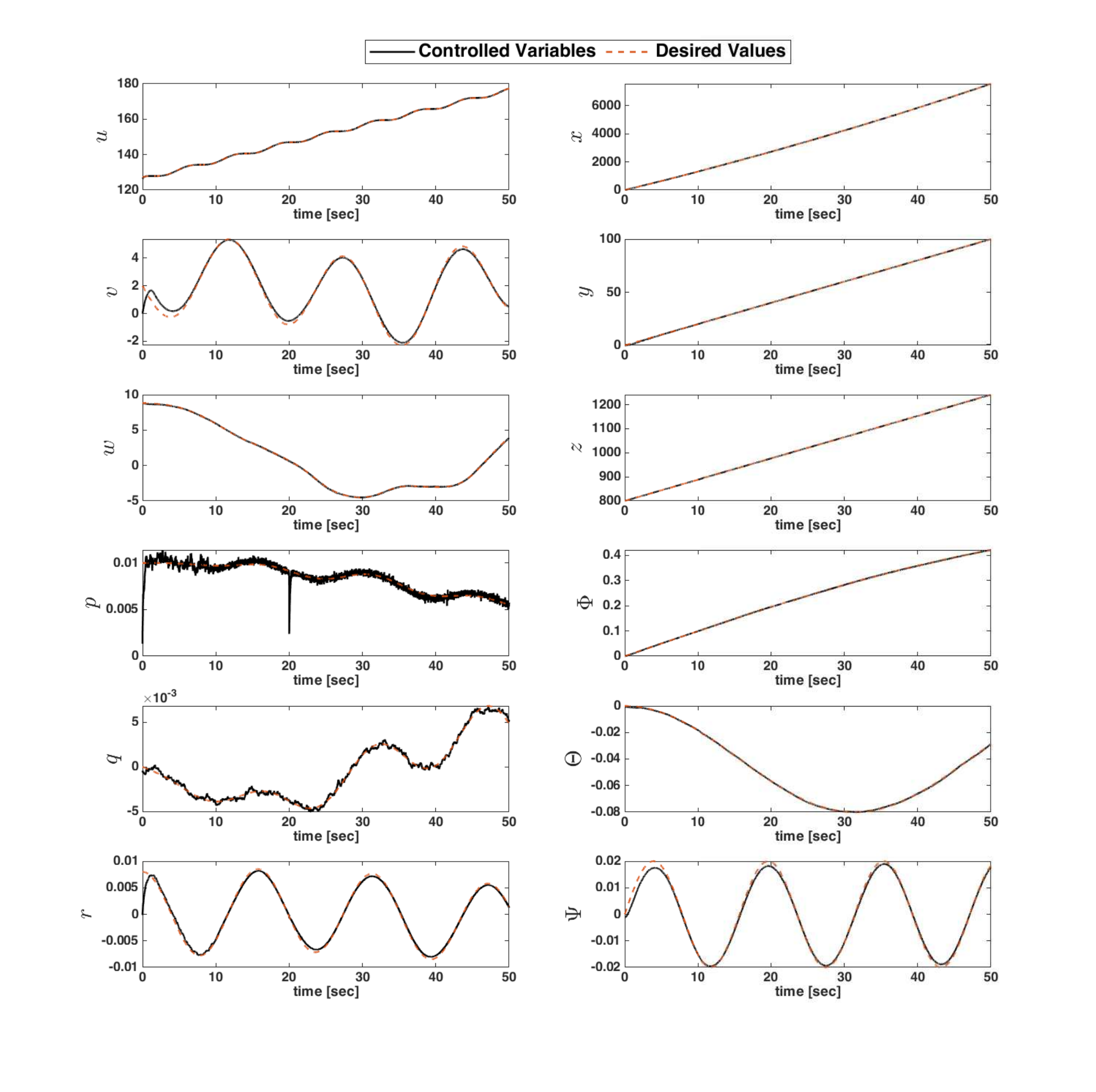}
  \caption{Tracking performance of the SDAC - SDAC is enabled at $t=10$ [sec] and uncertainty is happening at $t=20$} [sec]\label{fig:SDAC_states}
\end{figure}

The GTM underwent several closed-loop simulations with both the SDAC controller enabled and disabled in this section. The primary goal of the control was to track the reference trajectories and velocities, as shown in Figure \ref{fig:SDAC_states}. This figure illustrates the SDAC's tracking performance during a severe sudden uncertainty in $D$ and $B$ that occurred at $t=20$ seconds. To accurately capture the linear momentum dynamics, the SDAC's initiation time took 10 seconds, and it remained active onwards.

\begin{figure}[H]
  \centering
  \includegraphics[width=0.7\columnwidth]{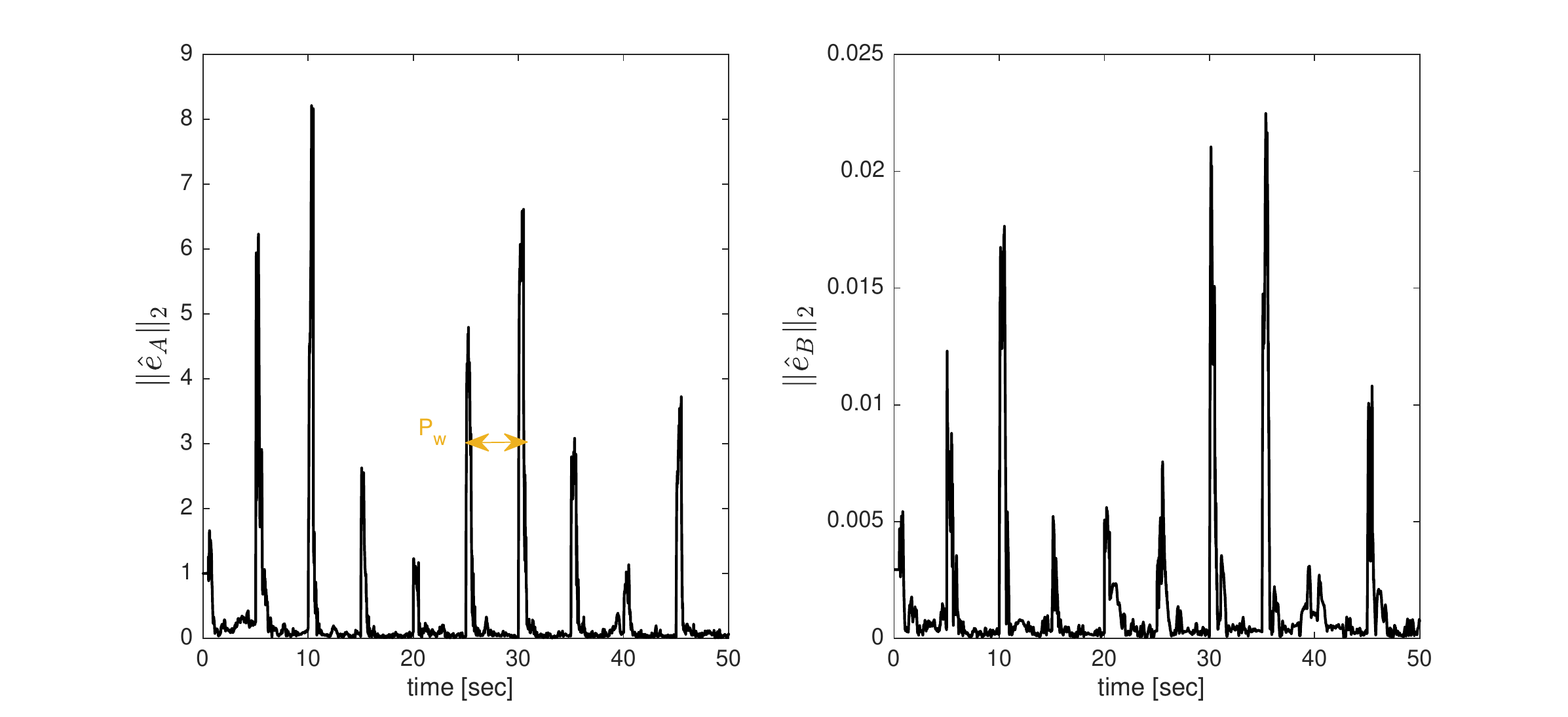}
  \caption{Estimation of momentum state transition matrices $(A,B)$ by Koopman operator and DMDc method}\label{fig:SDAC_e_AB}
\end{figure}

\begin{figure}[H]
  \centering
  \includegraphics[width=0.45\columnwidth]{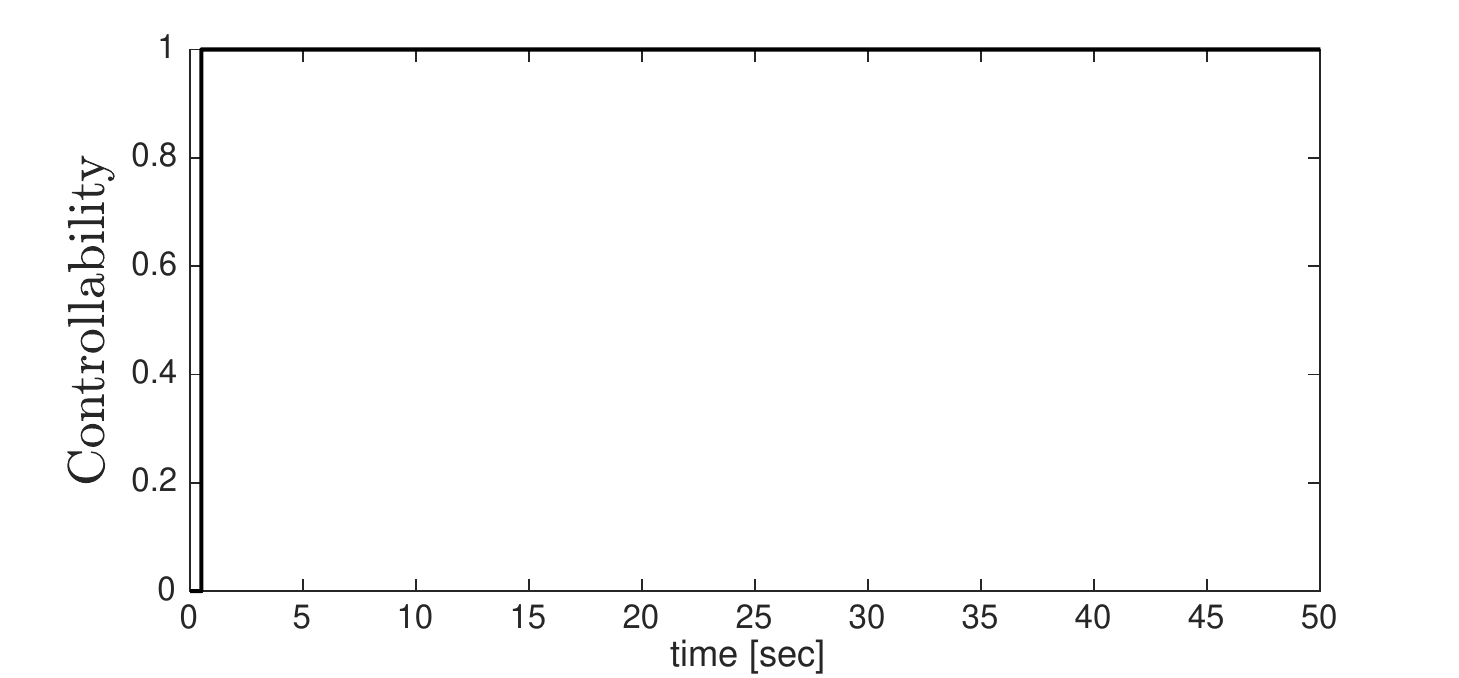}
  \caption{Controllability through time - value 1 means controllable}\label{fig:SDAC_ctrb}
\end{figure}

\begin{figure}[H]
  \centering
  \includegraphics[width=0.8\columnwidth]{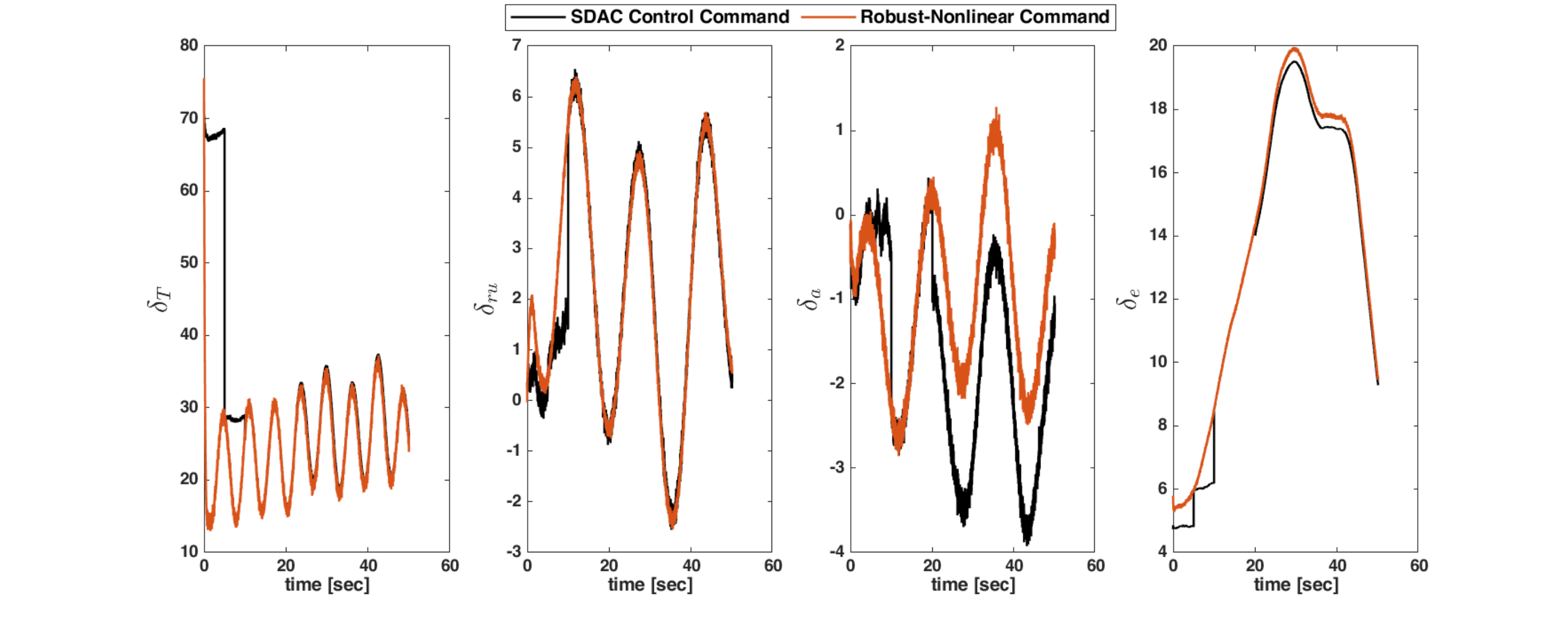}
  \caption{Control commands of SDAC and robust sliding mode controller}\label{fig:SDAC_delta}
\end{figure}

\begin{figure}[H]
  \centering
  \includegraphics[width=0.85\columnwidth]{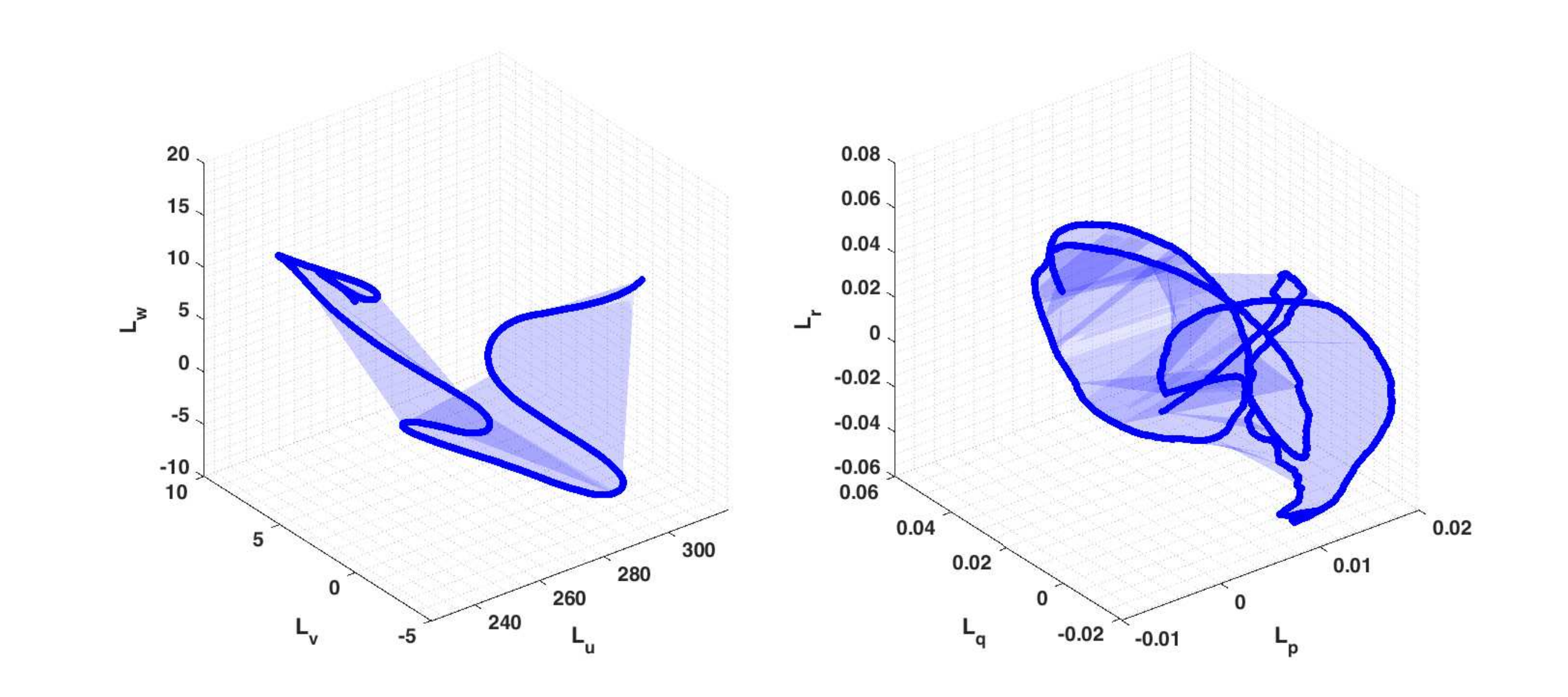}
  \caption{Reference momentum generated by the robust-nonlinear controller}\label{fig:SDAC_L_d}
\end{figure}

\begin{figure}[H]
  \centering
  \includegraphics[width=0.9\columnwidth]{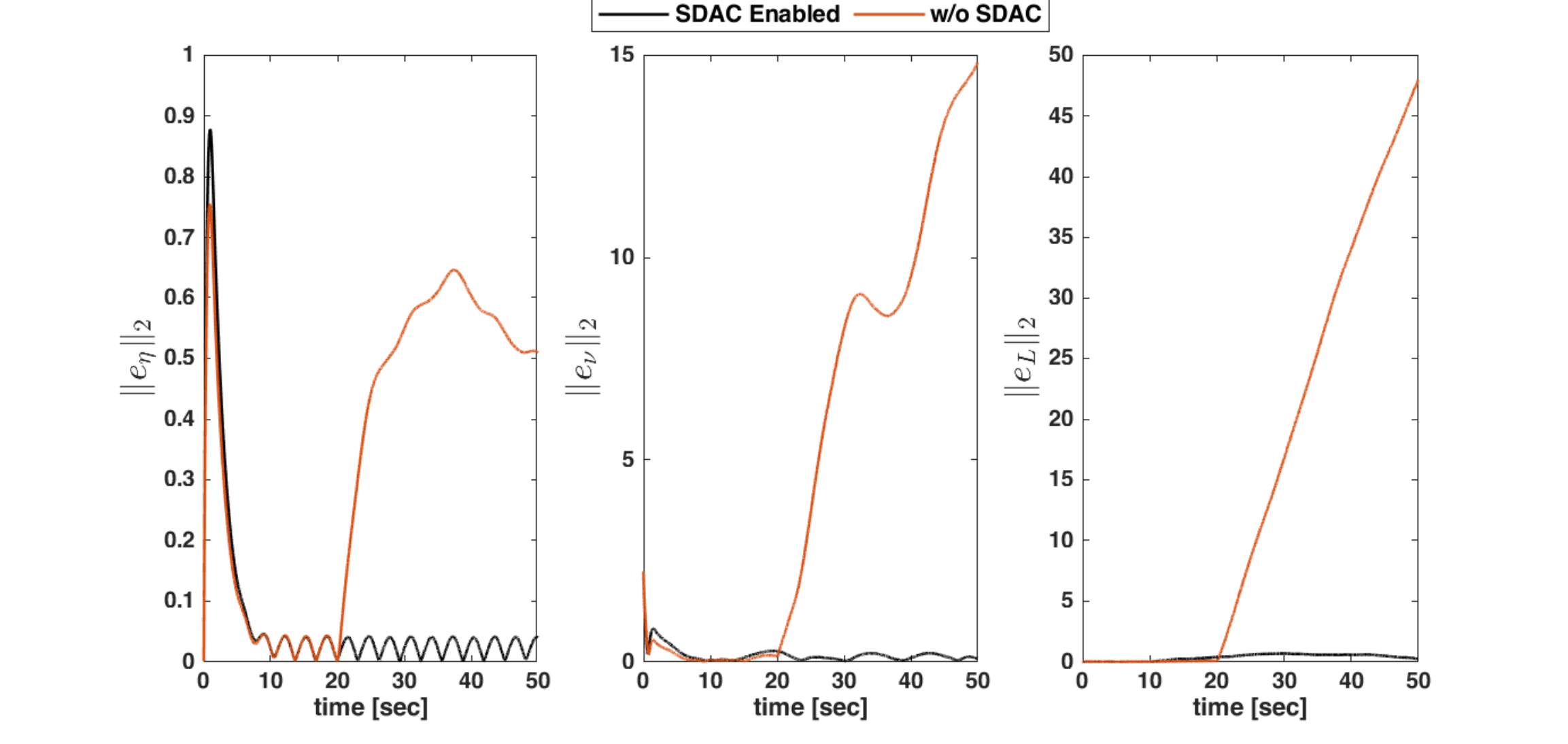}
  \caption{Earth-Body coordinates error and momentum tracking error}\label{fig:SDAC_error}
\end{figure}

In Figure \ref{fig:SDAC_e_AB}, the error of the Koopman operator for identifying momentum dynamics is demonstrated in comparison with the analytical equation in \eqref{equ:momentum}. The LQR controller gain is calculated at the end of each time window $P_w$ once estimation converges, and controllability of the couple $(A,B)$ is checked throughout the simulation, as shown in Figure \ref{fig:SDAC_ctrb}. The control effectors' values are depicted in Figure \ref{fig:SDAC_delta}, which shows that all control commands are within the admissible region. Consequently, the momentum subset in Figure \ref{fig:SDAC_L_d} is attainable, and flight is maneuverable for the simulated reference. In the initial phase ($t\leq 10$), the control commands of the SDAC and robust nonlinear controllers differ because the SDAC is inactive and there is no uncertainty. Therefore, the robust nonlinear controller tightly tracks the desired momentum, resulting in slight changes around the initial value of $\delta$ in state feedback at the beginning of the calculation window, $P_w$. Afterward, both controllers provide almost identical control commands until the introduction of sudden uncertainty at $t=20$ seconds, where differences become apparent. The error in tracking for body states, earth states, and momentum is shown in Figure \ref{fig:SDAC_error}, where the performance of the SDAC is superior compared with the robust nonlinear controller.


\section{Conclusion}
The design of a control system that combines sequential hybrid data-model techniques is investigated, with studying factors such as controllability stabilizability, robustness, and maneuverability during flight. Simulations have shown that utilizing data alongside the model can be beneficial in situations of uncertainty. The SDAC algorithm capitalizes on its lack of decision-making capabilities, allowing for independent tuning of reference momentum generation through the use of an internal dynamics robust nonlinear controller, as well as control of momentum dynamics derived by the Koopman operator DMDc. This independence enables the refinement of reference trajectories to achieve attainable momentum without affecting the momentum dynamics LQR control system.

\bibliography{SDAC_arxive}

\begin{thebibliography}{10}
\providecommand{\url}[1]{#1}
\csname url@samestyle\endcsname
\providecommand{\newblock}{\relax}
\providecommand{\bibinfo}[2]{#2}
\providecommand{\BIBentrySTDinterwordspacing}{\spaceskip=0pt\relax}
\providecommand{\BIBentryALTinterwordstretchfactor}{4}
\providecommand{\BIBentryALTinterwordspacing}{\spaceskip=\fontdimen2\font plus
\BIBentryALTinterwordstretchfactor\fontdimen3\font minus
  \fontdimen4\font\relax}
\providecommand{\BIBforeignlanguage}[2]{{%
\expandafter\ifx\csname l@#1\endcsname\relax
\typeout{** WARNING: IEEEtran.bst: No hyphenation pattern has been}%
\typeout{** loaded for the language `#1'. Using the pattern for}%
\typeout{** the default language instead.}%
\else
\language=\csname l@#1\endcsname
\fi
#2}}
\providecommand{\BIBdecl}{\relax}
\BIBdecl

\bibitem{eslami2023data}
M.~Eslami and A.~Banazadeh, ``Data-assisted control--a framework development by
  exploiting nasa gtm platform,'' \emph{arXiv preprint arXiv:2301.05646}, 2023.

\bibitem{jordan2004development}
T.~Jordan, W.~Langford, C.~Belcastro, J.~Foster, G.~Shah, G.~Howland, and
  R.~Kidd, ``Development of a dynamically scaled generic transport model
  testbed for flight research experiments,'' 2004.

\bibitem{bacon2007general}
B.~Bacon and I.~Gregory, ``General equations of motion for a damaged asymmetric
  aircraft,'' in \emph{AIAA atmospheric flight mechanics conference and
  exhibit}, 2007, p. 6306.

\bibitem{hueschen2011development}
R.~M. Hueschen, ``Development of the transport class model (tcm) aircraft
  simulation from a sub-scale generic transport model (gtm) simulation,'' 2011.

\bibitem{koopman1931hamiltonian}
B.~O. Koopman, ``Hamiltonian systems and transformation in hilbert space,''
  \emph{Proceedings of the national academy of sciences of the united states of
  america}, vol.~17, no.~5, p. 315, 1931.

\bibitem{hofmann2022advances}
C.~Hofmann, S.~Servadio, R.~Linares, F.~Topputo \emph{et~al.}, ``Advances in
  koopman operator theory for optimal control problems in space flight,'' in
  \emph{2022 AAS/AIAA Astrodynamics Specialist Conference}, 2022, pp. 1--13.

\bibitem{servadio2022dynamics}
S.~Servadio, D.~Arnas, and R.~Linares, ``Dynamics near the three-body libration
  points via koopman operator theory,'' \emph{Journal of Guidance, Control, and
  Dynamics}, vol.~45, no.~10, pp. 1800--1814, 2022.

\bibitem{arnas2021approximate}
D.~Arnas and R.~Linares, ``Approximate analytical solution to the zonal
  harmonics problem using koopman operator theory,'' \emph{Journal of Guidance,
  Control, and Dynamics}, vol.~44, no.~11, pp. 1909--1923, 2021.

\bibitem{durham2017aircraft}
W.~Durham, K.~A. Bordignon, and R.~Beck, \emph{Aircraft control
  allocation}.\hskip 1em plus 0.5em minus 0.4em\relax John Wiley \& Sons, 2017.

\bibitem{guo2011multivariable}
J.~Guo, G.~Tao, and Y.~Liu, ``Multivariable adaptive control of nasa generic
  transport aircraft model with damage,'' \emph{Journal of Guidance, Control,
  and Dynamics}, vol.~34, no.~5, pp. 1495--1506, 2011.

\bibitem{proctor2016dynamic}
J.~L. Proctor, S.~L. Brunton, and J.~N. Kutz, ``Dynamic mode decomposition with
  control,'' \emph{SIAM Journal on Applied Dynamical Systems}, vol.~15, no.~1,
  pp. 142--161, 2016.

\bibitem{brunton2016koopman}
S.~L. Brunton, B.~W. Brunton, J.~L. Proctor, and J.~N. Kutz, ``Koopman
  invariant subspaces and finite linear representations of nonlinear dynamical
  systems for control,'' \emph{PloS one}, vol.~11, no.~2, p. e0150171, 2016.

\bibitem{rowley2009spectral}
C.~W. Rowley, I.~Mezi{\'c}, S.~Bagheri, P.~Schlatter, and D.~S. Henningson,
  ``Spectral analysis of nonlinear flows,'' \emph{Journal of fluid mechanics},
  vol. 641, pp. 115--127, 2009.

\bibitem{kirk2004optimal}
D.~E. Kirk, \emph{Optimal control theory: an introduction}.\hskip 1em plus
  0.5em minus 0.4em\relax Courier Corporation, 2004.

\bibitem{gao.xue.2006}
D.~Gao and D.~Xue, ``Terminal sliding mode adaptive control for robotic
  manipulators,'' in \emph{Intelligent Control and Automation, 2006. WCICA
  2006. The Sixth World Congress on}, vol.~2, 2006, pp. 8853--8857.

\bibitem{garcia.salgado.torres.2009}
L.~Garcia-Valdovinos, T.~Salgado-Jimenez, and H.~Torres-Rodriguez, ``Model-free
  high order sliding mode control for rov: Station-keeping approach,'' in
  \emph{OCEANS 2009, MTS/IEEE Biloxi-Marine Technology for Our Future: Global
  and Local Challenges}, 2009, pp. 1--7.

\bibitem{uzmay.burkan.sarikaya.2004}
I.~Uzmay, R.~Burkan, and H.~Sarikaya, ``Application of robust and adaptive
  control techniques to cooperative manipulation,'' \emph{Control Engineering
  Practice}, vol.~12, no.~2, pp. 139--148, 2004.

\bibitem{eslami2019robust}
M.~Eslami, C.~S. Chin, and A.~Nobakhti, ``Robust modeling, sliding-mode
  controller, and simulation of an underactuated rov under parametric
  uncertainties and disturbances,'' \emph{Journal of marine science and
  application}, vol.~18, no.~2, pp. 213--227, 2019.

\bibitem{sato.tsurta.mukai.2007}
K.~Sato, K.~Tsuruta, and H.~Mukai, ``A robust adaptive control for robotic
  manipulator with input torque uncertainty,'' in \emph{SICE, 2007 Annual
  Conference}, 2007, pp. 1293--1298.

\bibitem{liu.goldenberg.1994}
G.~Liu and A.~Goldenberg, ``Asymptotically stable robust control of robot
  manipulators,'' in \emph{Robotics and Automation, 1994. Proceedings., 1994
  IEEE International Conference on}, vol.~4, may 1994, pp. 2968--2973.

\bibitem{effatnejad.namvar.2009}
K.~Effatnejad and M.~Namvar, ``Adaptive robust control of robot manipulators
  subject to input-dependent uncertainties,'' in \emph{Industrial Electronics
  and Applications, 2009. ICIEA 2009. 4th IEEE Conference on}, 2009, pp.
  3428--3433.

\bibitem{khalil.2002}
H.~Khalil, \emph{Nonlinear systems}.\hskip 1em plus 0.5em minus 0.4em\relax
  Prentice hall Upper Saddle River, 2002, vol.~3.

\bibitem{d1992reachability}
P.~d'Alessandro and E.~{De Santis}, ``Reachability in input constrained
  discrete-time linear systems,'' \emph{Automatica}, vol.~28, no.~1, pp.
  227--229, 1992.

\bibitem{alanwar2022robust}
A.~Alanwar, Y.~St{\"u}rz, and K.~H. Johansson, ``Robust data-driven predictive
  control using reachability analysis,'' \emph{European Journal of Control},
  vol.~68, p. 100666, 2022.

\bibitem{bak2021reachability}
S.~Bak, S.~Bogomolov, P.~S. Duggirala, A.~R. Gerlach, and K.~Potomkin,
  ``Reachability of black-box nonlinear systems after koopman operator
  linearization,'' \emph{IFAC-PapersOnLine}, vol.~54, no.~5, pp. 253--258,
  2021.

\bibitem{lee2022gpu}
Y.~Lee, M.~Cho, and K.-S. Kim, ``Gpu-parallelized iterative lqr with input
  constraints for fast collision avoidance of autonomous vehicles,'' in
  \emph{2022 IEEE/RSJ International Conference on Intelligent Robots and
  Systems (IROS)}.\hskip 1em plus 0.5em minus 0.4em\relax IEEE, 2022, pp.
  4797--4804.

\bibitem{mare2007solution}
J.~B. Mare and J.~A. De~Don{\'a}, ``Solution of the input-constrained lqr
  problem using dynamic programming,'' \emph{Systems \& control letters},
  vol.~56, no.~5, pp. 342--348, 2007.

\bibitem{johansen2000explicit}
T.~A. Johansen, I.~Petersen, and O.~Slupphaug, ``On explicit suboptimal lqr
  with state and input constraints,'' in \emph{Proceedings of the 39th IEEE
  Conference on Decision and Control (Cat. No. 00CH37187)}, vol.~1.\hskip 1em
  plus 0.5em minus 0.4em\relax IEEE, 2000, pp. 662--667.

\bibitem{scokaert1998constrained}
P.~O. Scokaert and J.~B. Rawlings, ``Constrained linear quadratic regulation,''
  \emph{IEEE Transactions on automatic control}, vol.~43, no.~8, pp.
  1163--1169, 1998.

\end{thebibliography}
\bibliographystyle{IEEEtran}
\end{document}